# Development of high-level applications for High Energy Photon Source booster

Yuemei Peng,[1] Daheng Ji,[1] Hongfei Ji,[1] Nan Li,[1] Xiaohan Lu,[1] Saike Tian,[1] Yuanyuan Wei,[1] Haisheng Xu,[1] Yaliang Zhao,[1] Yi Jiao,[1,2*] Jingyi Li,[1]

[1] *Key Laboratory of Particle Acceleration Physics and Technology, Institute of High Energy Physics, Chinese Academy of Sciences,19BYuquan Road, Shijingshan District, Beijing 100049, China.*
[2] *University of the Chinese Academy of Sciences, 19A Yuquan Road, Shijingshan District, Beijing 100049, China.*
*Corresponding author(s)：Jiao Yi，*E-mail(s):* jiaoyi@ihep.ac.cn
*Corresponding author(s)：Xiaohan Lu，*E-mail*: Luxh@ihep.ac.cn

ABSTRACT: The High Energy Photon Source (HEPS), is the first fourth-generation storage ring light source being built in the suburb of Beijing, China. The storage ring was designed with the emittance lower than 60 pm.rad with a circumference of 1.36 km and beam energy of 6 GeV. Its injector contains a 500 MeV S-band Linac and a 454 m booster which was designed as an accumulator at the extraction energy. In the energy ramping control design of HEPS booster, the ramping process was programed to be able to stop and stay at any energy between the injection energy and the extraction energy. This feature enables us to conduct energy-dependent machine studies and ramping curve optimization. The beam commissioning of HEPS Linac finished in June, 2023. And the beam commissioning of booster started in the end of July, 2023. In November 17, main target values proposed in the preliminary design report has been reached. The high-level applications (HLAs) are essential tools for beam commissioning. The development of HLAs, which are based on the framework named Python accelerator physics application set (*Pyapas*), started in the end of 2021. The HEPS physics team spent more than one year to develop and test the HLAs to meet the requirements of beam commissioning of the booster. Thanks to the modular design, the principle based on physical quantities, and the ability of running simulation models online from the *Pyapas*, the development efficiency and reliability of the HLAs have been greatly improved. In particular, the principle based on physical quantities allows us to control the beam more intuitively.

# 1. Introduction

The High Energy Photon Source (HEPS) is an ultra-low emittance light source being built in the suburb of Beijing, China [1,2]. The facility is comprised of a full energy injector, a 6 GeV storage ring [3] with an emittance less than 60 pm.rad, more than ten beam lines and experiment stations. HEPS storage ring adopts the hybrid 7-bend achromat approach [4] with special magnets, such as high gradient quadrupoles, longitudinal gradient dipoles [5,6], combined-function magnets with transverse gradients and anti-bends [7], so as to achieve an ultra-low emittance. On-axis swap-out injection scheme [8] has been proposed for the HEPS storage ring associated with the limited dynamic aperture, and it requires a full energy injector to provide full bunch charge in once injection. The HEPS injector is composed of a 500 MeV S-band Linac [9], a full energy booster [10] which can be used as an accumulation ring at 6GeV, and three transfer lines [11] connecting the Linac, booster and storage ring. The beam commissioning of Linac and booster has been finished in 2023 and the storage ring commissioning is scheduled in the middle of 2024.

To meet the demand for the storage ring injection charge, booster "high energy accumulation" scheme was chosen. The booster should install a high energy injection system and have the stability to ramping a bunch charge more than 5 nC to extraction energy. So, the HEPS booster was designed a four-fold symmetry lattice with a circumference of approximately 454 m, about 1/3 of that of the storage ring. Each period is composed of 14 identical cells together with two matching cells to the dispersion free straight sections. The physics design of the booster was finalized at the end of 2019, with the layout and optics of one super-period presented in Fig.1.

The milestones of the booster installation and commissioning include: start of the pre-alignment installation of magnet units on March 1, 2022; start of the tunnel installation of magnet units in the booster on August 8, 2022; completion of the full-ring vacuum connection of the booster on January 13, 2023; start of beam commissioning in the booster on July 25, 2023; first ramping the beam energy to 6 GeV on August 9, 2023; successful ramping the beam energy of an electron beam with a single bunch charge greater than 5 nC to 6 GeV on November 6, 2023; and main parameters of the HEPS booster successfully achieved their target values on November 17, 2023[12].

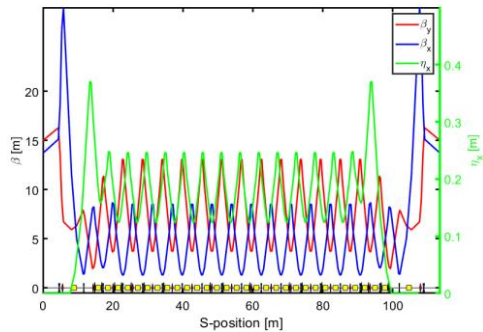

Fig. 1. Layout and optics of a super-period in the booster.

As mentioned above, the main challenge of the booster design and commissioning is ramping a bunch charge more than 5 nC to 6 GeV with a high transmission efficiency, and the designed value of the transmission efficiency during injection and ramping process in the HEPS booster is not less than 80% [12]. In order to achieve such high transmission efficiency, we need to measure and make necessary corrections to parameters in the booster, such as the beam orbit, tune, dispersion, chromaticity, beam lifetime, and so on. Many efficient high-level applications (HLAs) are essential for commissioning and operation of the HEPS booster. HLAs should provide real-time data monitoring, automatic adjustment, and optimization functions, helping accelerator operators quickly and accurately adjust beam parameters to achieve the desired accelerator performance.

The HLAs for the HEPS booster commissioning contains control applications, measurement applications and monitor applications. To provide the environment for HLA development, a virtual accelerator with channel access server has been implemented to emulate the real machine. The virtual accelerator is based on the independently developed application framework named Python accelerator physics application set (Pyapas) [13,14,15],which chooses Python as the main developing programming language. Pyapas is designed with a modular concept, and can be divided to several modules: graphic operator interface (GUI) module, dual-layer physical module, communication module, client-server module, database module, and pre-development module. Each module is designed individually according to its functionality, ensuring its completeness and independence [13]. For a quick access to frequently used applications, all of the HLAs of HEPS are organized as a list in the launcher which is similar to a file browser.

The development of HEPS booster HLAs strated in the end of 2021 when the developing and testing of HLAs for the Linac was finished [16]. The accelerator physics team spent more than one year for developing and testing the HLAs based the virtual accelerator to meet the requirements of the booster commisiioning. For the ease of team collaborations, version control system git [17] is utilized to manage the source code. During the beam commissioning process, we further modified and improved the HLAs according to actual usage requirements.

In this paper we will introduce the HLAs developed for HEPS booster commissioning and future operation. The algorithm, logic implementation and GUI of all the HLAs used in the HEPS booster will be described in Section 2. In section 3, we will briefly introduce the application of HLAs to the beam commissioning for monitoring and controlling various parameters of the accelerator. In the section 4 we give a brief summary.

## 2. THE GUI AND ALGORITHM OF APPLICATIONS

The control applications of HEPS booster can directly controls the physical quantities of the accelerator elements, such as the bending angle, beam energy, and so on. The control applications used in HEPS booster include booster controller, first turn commissioning, global orbit correction, local orbit correction and tune adjusting correction. The measurement applications are used to measure the beam parameters, such as beam emittance and energy spread, dispersion, chromaticity, transmission efficiency, and so on. Due to the fact that the quadrupole and sextupole magnets are powered in series in the HEPS booster, it was decided not to perform beam-based alignment. In this section, the applications used in the HEPS booster will be introduced one by one.

### 2.1 Booster control application

As a full energy injector, the HEPS booster increases the beam energy from 500 MeV to 6 GeV, the same as that of the storage ring. The full-energy beam could be injected directly to the storage ring or after merged with bunches re-injected from the storage ring by using a "high-energy accumulation" scheme [12]. For more flexible beam commissioning, the power supplies and RF cavities were designed with two control modes: ramping mode and tuning mode. These two modes can be switched online. The ramping mode is used for beam energy ramping, and the tuning mode is used for beam parameter optimization at any energy point. In the latter mode, the magnet power supplies can remain at any current point, staying in DC mode for optimization.

To address the swift optimization and adjustment requirements of booster magnet settings, the booster control application was developed, as shown in Fig. 2.

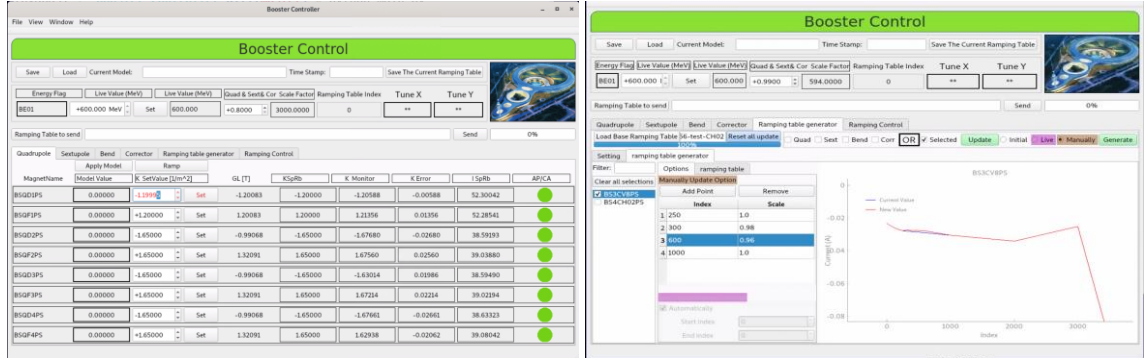

**Fig. 2.** The booster control application

Booster control application allows operators to set or adjust current values of each magnet power supply, based on physical quantities in tuning mode. After completing the optimization of these current values at different energies, the current values of the magnet power supplies' ramping curve for a specific energy can be updated online. This ability of point-by-point optimization and online updating of the ramping curve significantly enhances the efficiency of beam commissioning. Booster control application offers multiple features for updating the ramping curve online. For instance, it can generate the complete ramping curve for the whole period, relying on the optimized physical quantities of the magnet at any energy point. Especially, it allows proportional adjustment of a segment of magnet power supply ramping curve, while keeping the current set values at other energy points unchanged.

Additionally, the booster control application includes a save-restore function. This allows the operator to save all magnet settings at a specific energy point to a file and restore them to the machine when needed. The ramping table for all AC magnets can also be saved to a folder and restored to the machine as needed. The folder and file name will be the model name. Currently, the save and restore strategy is relatively simple. A more robust application for saving and restoring settings for the entire machine is under development.

## 2.2 The turn-by-turn (TBT) data display application

The TBT Data display application is utilized for the display of TBT data during the initial injection phase of commissioning and short-term analysis both in the HEPS booster and storage ring. This application can obtain data of 128 turns using waveform format PV channels and processing TBT data of much more than 128 turns from offline data files. It is designed to effectively handle TBT data containing noise. The GUI of this application was shown in Fig. 3.

– 3 –

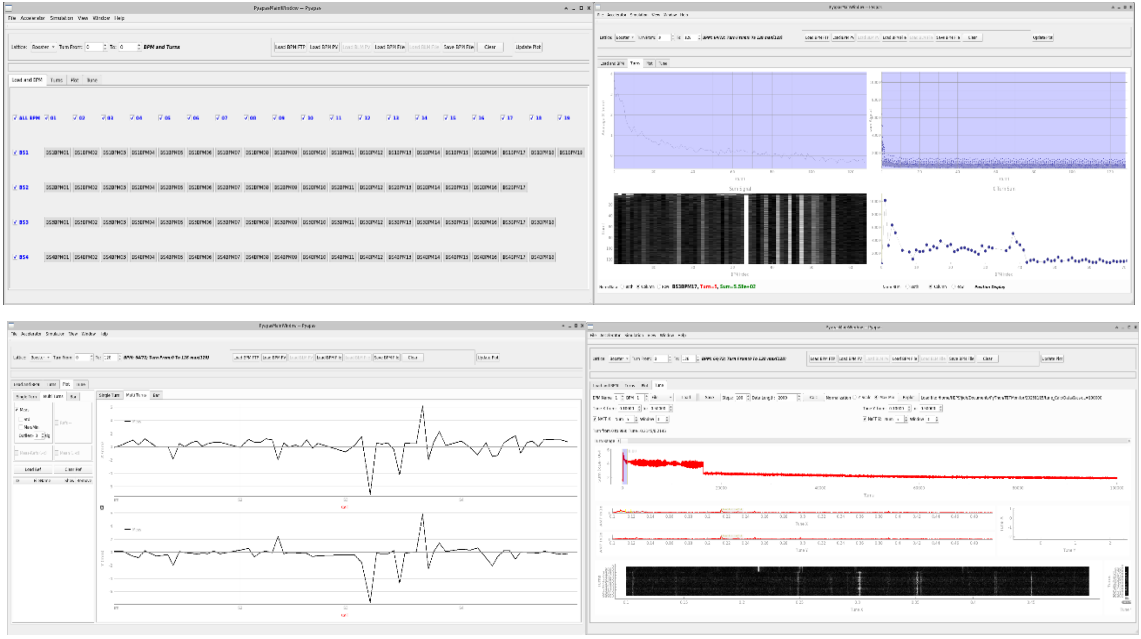

**Fig. 3.** The GUI of TBT data display application

The TBT data display application is composed of four Tabs. The first tab named "Load and BPM", is used for choosing the machine and the BPM to be used. The second tab named "Turns" is used to evaluate the performance of each BPM signal and the effective number of turns for beam transmission. It is divided into four sub-images, the upper-left image displays the average values of beam horizontal trajectory of each turn given by all BPMs, reflecting the beam energy change. The upper-right graph displays the average values of sum signal given by all BPMs for each turn, which can reflect the bunch charge change versus turn index based on the fact that the BPM sum signals are proportional to the beam current. The lower-left graph shows a waterfall plot of TBT sum signal with normalized amplitude, reflecting losses during beam transmission. The lower-right graph displays beam loss data per turn, with a display format and normalization settings similar to the lower-left graph. "Plot" is the third tab, which is used for displaying the beam trajectory and the STD values of each BPM data. The reference orbit, difference between the measured orbit and reference orbit are shown in this tab. The fourth tab is "Tune", which shows the tune obtained through numerical analysis of the fundamental frequencies of thousands' turns of TBT data.

### 2.3 Global orbit correction application

The alignment and field errors of various magnets will cause the real orbit and optics parameters are different from the theoretical values. The parameter variation will affect the performance of the booster. In order to eliminate this effect, we perform orbit correction based on response matrix. Changes in the strength of correctors can lead to changes in the particle orbit. The relationship between the change of particle orbit $\Delta u_{(s)}$ and the corrector strength change $\Delta\theta_m$ in $x$ or $y$ direction is shown in Eq. (2.3.1).

$$\Delta u_{(s)_n} = R_{nm}\, \Delta\theta_m \tag{2.3.1}$$

where R is the response matrix, *n* is the number of used BPMs, and *m* is the number of correctors.

To keep the path length to be close to the designed circumference in the process of correction, it is better to make the sum of correctors' strength close to zero,

$$sum \sum_{i=1}^{m} \Delta\theta_i = 0 \tag{2.3.2}$$

The relationship of Eq. (2.3.1) can be presented as

$$\begin{pmatrix} x_1 \\ x_2 \\ \vdots \\ x_n \\ 0 \end{pmatrix} = \begin{pmatrix} R_{11} & \cdots & R_{1m} \\ \vdots & \ddots & \vdots \\ R_{n1} & \cdots & R_{nm} \\ w_1 & \cdots & w_1 \end{pmatrix} \begin{pmatrix} \Delta\theta_1 \\ \vdots \\ \Delta\theta_m \end{pmatrix} \tag{2.3.3}$$

where $w_1$ is the weights imposed to correctors' strengths, $w_1 \neq 0$.

In ring accelerators, the strength change of horizontal correctors will induce beam energy change, the relationship between the energy and horizontal corrector's strength is shown in Eq. (2.3.4), where $D_{x,i}$ is the horizontal dispersion function at the $i_{th}$ corrector, $\alpha_p$ is the momentum compaction factor and C is the circumference of the ring.

$$\frac{\Delta E_i}{E} = \frac{\Delta\theta_{x,i} D_{x,i}}{\alpha_p C} \tag{2.3.4}$$

To enhance the versatility, dispersion correction and energy correction also be added in the algorithm of horizontal orbit correction. Assuming the weight of dispersion correction and energy correction are $w_2$ and $w_3$, respectively, the relationship of orbit correction with dispersion correction could be shown as Eq. (2.3.5), and the relationship of orbit correction with energy correction is presented as Eq. (2.3.6).

$$\begin{pmatrix} x_1 \\ x_2 \\ \vdots \\ x_n \\ w_2 \Delta D_1 \\ w_2 \Delta D_2 \\ \vdots \\ w_2 \Delta D_n \end{pmatrix} = \begin{pmatrix} R_{11} & \cdots & R_{1m} \\ \vdots & \ddots & \vdots \\ R_{n1} & \cdots & R_{nm} \\ w_2 M_{11} & \cdots & w_2 M_{1m} \\ \vdots & \ddots & \vdots \\ w_2 M_{n1} & \cdots & w_2 M_{nm} \end{pmatrix} \begin{pmatrix} \Delta\theta_1 \\ \vdots \\ \Delta\theta_m \end{pmatrix} \tag{2.3.5}$$

where $M$ is the response of dispersion function to corrector strength.

$$\begin{pmatrix} x_1 \\ x_2 \\ \vdots \\ x_n \\ w_3 \frac{\Delta E}{E} \end{pmatrix} = \begin{pmatrix} R_{11} & \cdots & R_{1m} \\ \vdots & \ddots & \vdots \\ R_{n1} & \cdots & R_{nm} \\ w_3 \frac{D_{x,1}}{\alpha_c C} & \cdots & w_3 \frac{D_{x,m}}{\alpha_c C} \end{pmatrix} \begin{pmatrix} \Delta\theta_1 \\ \vdots \\ \Delta\theta_m \end{pmatrix} \tag{2.3.6}$$

This application consists of two relatively independent modules, one is for measuring the response matrix, and the other is for orbit correction. The layout of response matrix measurement module is shown in Fig. 4. Operators can select BPMs and correctors used for measurement. If the beam current is below a certain value, the measured response will have a large error, so the threshold of beam current should be set to ensure the validity. When the beam current is lower than the limit in measurement process, the measure will be paused, and a reminding message will be given. Operators can also adjust the corrector strengths in each step and choose steps of one corrector to be used for reducing the errors during the measurement process.

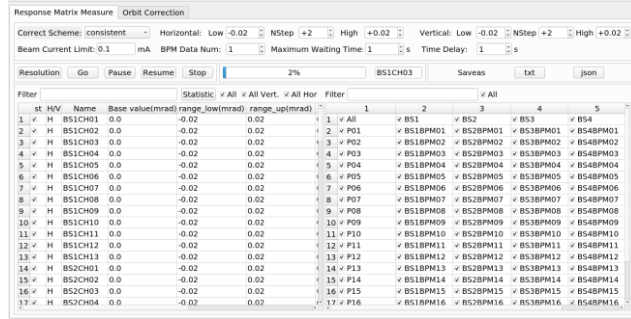

**Fig. 4.** The GUI of response matrix measurement module

The orbit correction module, as shown in Fig. 5, includes four sub-pages. The "orbit" page is used for orbit display, the "setting" page is designed to choose the correctors and BPMs used for orbit correction, and the "SVD" page is used for displaying singular values and their threshold, and the "info" page is a spare page, where additional information can be added if needed. The process of orbit correction will start once the button 'calc' is clicked. The response matrix used in orbit correction can be measured online or loaded from files. The correction target could be set as "zero orbit" or "golden orbit" that loaded from a file. The functions of saving orbit and comparing with history setting are also implemented in this module.

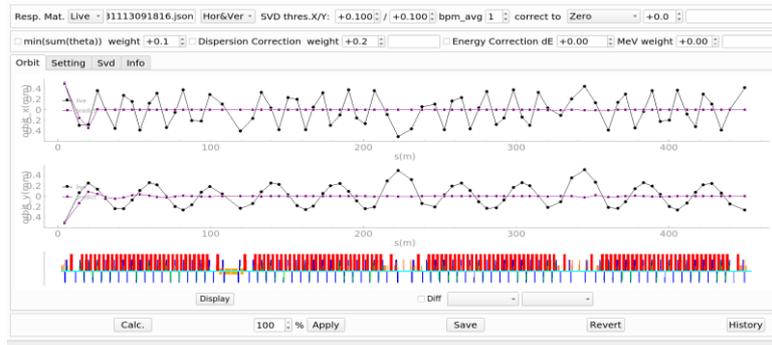

**Fig. 5**. The GUI of global orbit correction module

### 2.4 Local orbit correction application

Local orbit correction application is used to make a local bump, which can control the height or angle of the local orbit, without affecting the orbit out of this region, especially in the beam injection and extraction region.

To control the orbit more exactly, the response matrix method is adopted in this application. The orbit distortion $\Delta u$ induced by the 3-corrector bump at the position of BPMs can be expressed as

$$\Delta u = (0_{BPM1} \cdots 0_{BPMu}\ h_{BPMc}\ 0_{BPMd} \cdots 0_{BPMm})' \qquad (2.4.1)$$

where the subscript *BPMu* means the first upstream BPM outside of the local region, *BPMd* means the first downstream BPM outside of the region, *BPM1* is the first BPM in the lattice, and *BPMm* is the last BPM in the lattice. Since it is supposed the local bump does not have any effect on the orbit outside the bump region, the amount of orbit change is limited to zero at BPMs outside the bump region. *BPMc* is the BPM inside the 3-bump region which is used to assign the height of the local bump. For other BPMs inside the bump, the orbit distortions are out of control. $h_{BPMc}$ represents the height of the local bump.

The SVD decomposition of the response matrix is performed, and then the strength changes correlated to the bump height $h_{BPMc}$ can be solved according to Eq. (2.4.2).

$$R = USV^T \qquad (2.4.2)$$

where $U$ and $V$ are orthogonal matrices and $S$ is the diagonal matrix of singular values.

When 4-corrector bump is applied, it is possible to control the angle of the bump as well. For the bump height adjustment, the vector $\Delta u$ is expressed as:

$$\Delta u = (0_{BPM1} \cdots 0_{BPMu}\ h_{BPMc1}\ h_{BPMc2}\ 0_{BPMd} \cdots 0_{BPMm})' \qquad (2.4.3)$$

where BPMc1 and BPMc2 are the two BPMs assigned in the bump area. For the angle adjustment, the vector $\Delta z$ is expressed as:

$$\Delta z = (0_{BPM1} \cdots 0_{BPMu}\ -angle*\frac{L}{2}_{BPMc1}\ angle*\frac{L}{2}_{BPMc2}\ 0_{BPMd} \cdots 0_{BPMm})' \qquad (2.4.4)$$

where L is the distance between the BPMc1 and the BPMc2, and *angle* is the amount of angle adjustment. Similar to the Eq. (2.4.2), strength changes of 4-corrector bump can be obtained for the amount of height adjustment or angle adjustment, respectively.

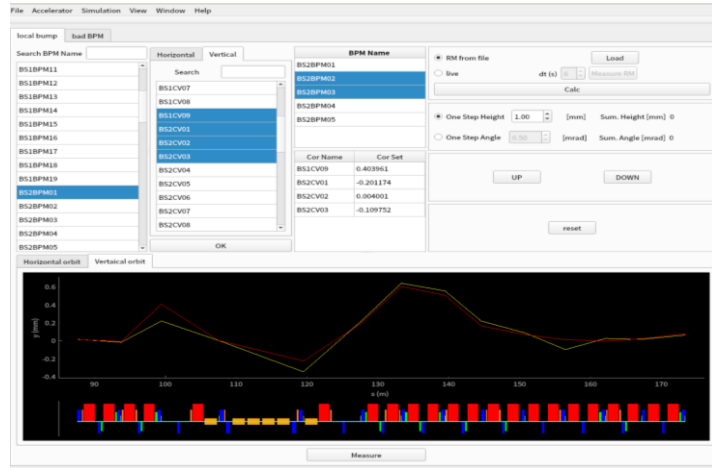

**Fig. 6.** The GUI of the local bump application.

Fig. 6 shows the GUI of the local bump application. It consists of bump configuration, bump adjustment and the graphic display. In the bump configuration, the correctors and BPMs to be used should be chosen first. In the adjustment area, response matrix can be loaded from file or measured online. The bump adjustment is carried out by clicking the "up" or "down" buttons to increase or decrease the bump. The local orbit distortions after adjustment are then displayed in the graphic area.

**2.5 Tune adjusting application**

The quadrupole magnets in the HEPS booster are divided into 8 groups, with each group powered by a single power supply. The tune of HEPS booster is corrected by adjusting the strengths of two groups of quadrupole magnets, where one is horizontal focusing (QF04) and the other is horizontal defocusing (QD04). For a small tune variation $\Delta\upsilon$, the relationship between $\Delta\upsilon$ and the strength variation of magnets can be expressed as a response matrix shown in Eq.(2.5.1)

$$\begin{pmatrix}\Delta v_x \\ \Delta v_y\end{pmatrix} = \begin{pmatrix} \dfrac{\partial v_x}{\partial K_{QF}} & \dfrac{\partial v_x}{\partial K_{QD}} \\ \dfrac{\partial v_y}{\partial K_{QF}} & \dfrac{\partial v_y}{\partial K_{QD}} \end{pmatrix} \begin{pmatrix}\Delta K_{QF} \\ \Delta K_{QD}\end{pmatrix}$$

（2.5.1）

where $K_{QF}$ and $K_{QD}$ are the strengths of QF04 and QD04, respectively.

The default response matrix is calculated based on the bare lattice, using the OCELOT simulation code [18]. However, this option can be manually changed by the operator. If the PVs of tune measurement results are available, this program can also calculate the practical response matrix by varying $K_{QF}$ and $K_{QD}$.

To test the validity of this response matrix method, we compare the tune from response matrix with that from the OCELOT simulation, which are shown in Fig. 7. We can see that for a 1.5% variation in quadrupole strength, the disparity between the simulation and matrix methods is less than 0.006, which is a small error.

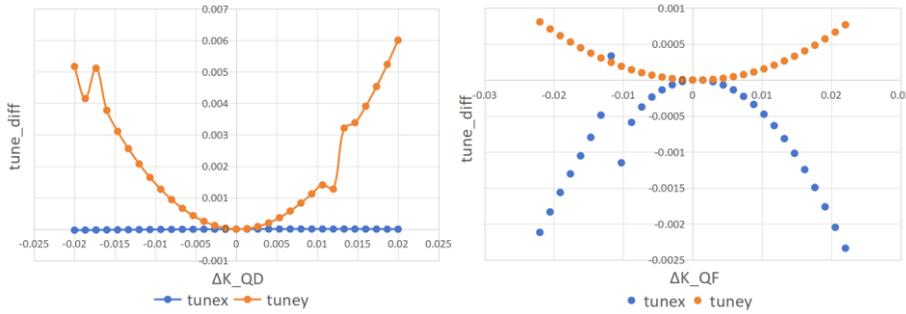

**Fig. 7.** The tune disparity between the simulation and response matrix methods, for different quadrupole strengths

The GUI of the application shows both the measured value of tunes (labelled as "live") and the theoretical tune value (labelled as "simulator"). Several methods are available for this application: correcting the tune based on the "live" value, the "simulator" value, or inputting the $\Delta v$ value directly. The theoretical beta-beating corresponding to the design value is also plotted on the panel, which is obtained from the OCELOT simulation.

The GUI of this application is shown in Fig. 8, and the usage is simple:
(1) The operator inputs the target tune or $\Delta v$
(2) Click the "calculate" button to calculate $K_{QF}$ and $K_{QD}$
(3) Click the "ramp" button to apply the change of the strength of magnets

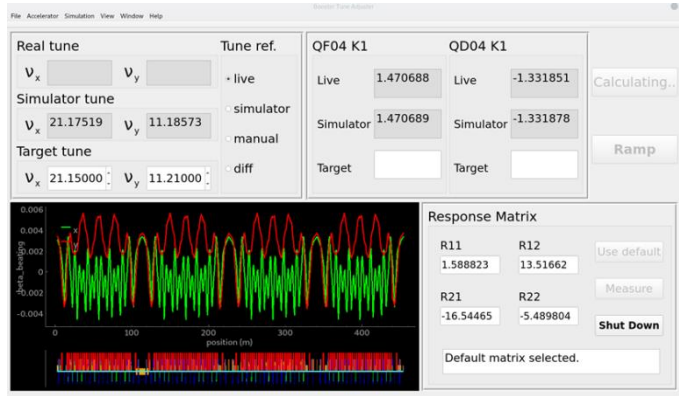

**Fig. 8.** The GUI of the tune adjusting application.

## 2.6 Emittance and energy spread measurement application

There are two synchrotron light beamlines for measuring the emittance and energy spread respectively in HEPS booster. The beamline for measuring emittance is located at the bending magnet BS1B01, which is at a nearly dispersion-free region. The beamline for measuring energy spread is located at the bending magnet BS1B06, where the horizontal dispersion has a finite value. The GUI of this application is shown in Fig. 9.

The synchrotron light is monitored by the CCD camera and then digitized. The generated matrix is put into the PV channel and monitored by the application. The horizontal and vertical beam size are then obtained by projecting the data to *x* and *y* axes, respectively. The beam size values are then used to calculate the emittance and the energy spread from Eqs. (2.6.1) and (2.6.2), respectively. The obtained emittance will be used in the measurement of energy spread.

$$\varepsilon_{x,y} = \sigma_{x,y}^2/\beta_{x,y} \tag{2.6.1}$$

$$\sigma_\delta = \frac{\sqrt{\sigma_x^2 - \beta_x \varepsilon_x}}{D_x} \tag{2.6.2}$$

where the $\sigma_{x,y}$ is the measured bunch size, $\beta_{x,y}$ is the beta function, $\varepsilon_{x,y}$ is the measured horizontal emittance, $D_x$ is the dispersion, $\sigma_\delta$ is the energy spread.

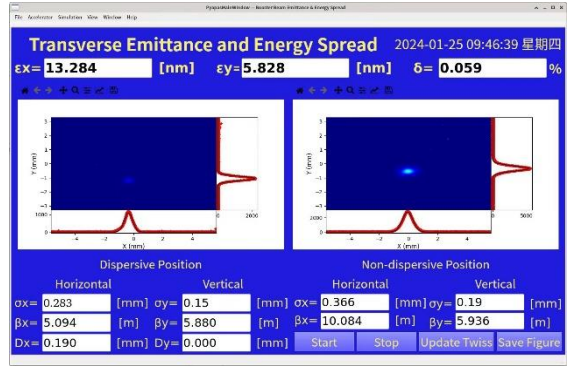

**Fig. 9.** GUI of the application for measurement of the transverse emittance and energy spread.

## 2.7 Dispersion measurement application

The dispersion function characterizes the transverse position deviation of particles with respect to their relative energy deviation when they traverse a magnetic or electric field [19]. The transverse position of an individual particle can be described by Eq. (2.7.1):

$$u(s) = u_{c.o.}(s) + u_\beta(s) + D(s)\frac{\Delta p}{p} \tag{2.7.1}$$

where $u_{c.o.}(s)$ represents the closed orbit, $u_\beta(s)$ represents the orbit deviation caused by betatron oscillation, $D(s)\frac{\Delta p}{p}$ represents the orbit deviation caused by energy deviation, $D(s)$ is the dispersion function, and $\frac{\Delta p}{p}$ represents the particle's momentum deviation relative to the reference particle.

The dispersion function can be obtained through calculating the energy shift with the RF frequency change. During the measurement process, it is required to maintain beam stability within the frequency adjustment range to avoid beam loss.

By varying the RF frequency and acquiring the data from BPMs, the orbit deviation caused by the change of beam energy can be obtained. As shown in Eq. (2.7.2), the variation of beam orbit to the RF frequency's change one can obtain a slope that contains information about the dispersion function.

$$\Delta u(s) = D(s)\frac{\Delta p}{p} = \frac{D(s)}{\alpha_p}\frac{\Delta C}{C} = -\frac{D(s)}{\eta}\frac{\Delta f_{rf}}{f_{rf}} \qquad (2.7.2)$$

where C represents the circumference, substituted with the theoretical design value, and $\eta = \alpha_p - \frac{1}{\gamma^2}$.

Fig. 10 shows the GUI of this application, which allows operators to manually or automatically adjust the RF frequency for dispersion measurement. One can select valid frequency measurement points and view the BPM data.

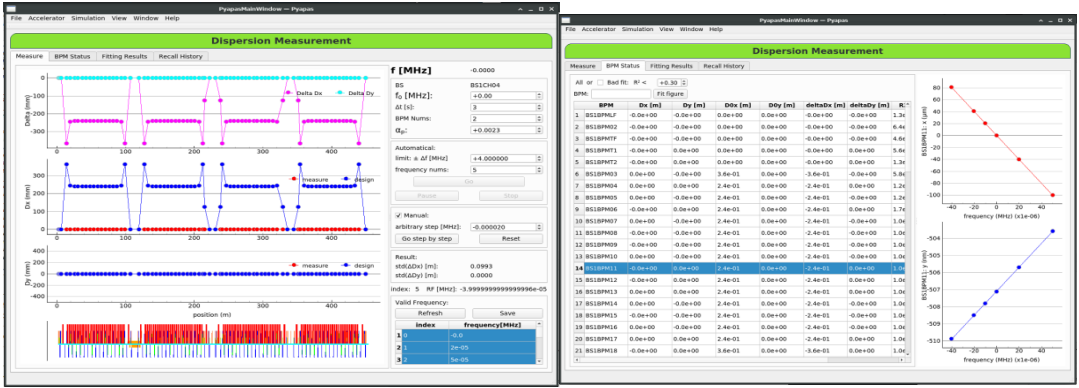

(a) the main GUI (b) the raw BPM data and the fitting result

**Fig. 10.** The GUI of dispersion function measurement application.

## 2.8 Chromaticity measurement application

The measurement method for chromaticity applied here is widely used in various accelerators, i.e., measuring the tune shifts due to the energy variation by changing the RF frequency. The relationship is given in (2.8.1).

$$\Delta \nu = -\xi \cdot \frac{1}{\alpha_p} \cdot \frac{\Delta f_{rf}}{f_{rf}} \qquad (2.8.1)$$

where $-\xi \cdot \frac{1}{\alpha_p}$ is a slope of the tune deviation with the RF frequency tuning. During the measurement, the RF frequency is gradually changed in a limited scope, which is determined by the energy acceptance. The tunes are sampled after each tuning. The measurement data are fitted linearly, and the fitting slope can be derived.

The GUI of the chromaticity measurement application is shown in Fig. 11. The scope of RF frequency tuning and the number of sampling need to be configured before the start of measurement. For the way of measurement, automatic measurement and manual measurement are two options. When the measurement is completed, the measure data is plotted together with fitting curve, and the chromaticities for both planes are displayed. Extra data can be added by carrying out individual measurement. The abnormal data can be dropped and the results will be refreshed.

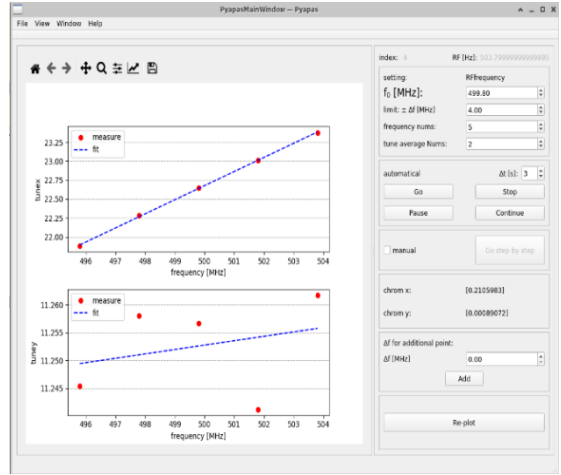

**Fig. 11.** The GUI of the chromaticity measurement application

**2.9 Injection efficiency and beam lifetime measurement application**

For a booster, the beam lifetime is not a major concern due to the relatively short cycle time. However, as the HEPS booster can be operated at any energy from 500 MeV to 6 GeV like a storage ring, the beam lifetime measurements at each energy point is considered in the HEPS booster commissioning. The beam lifetime, injection efficiency and transmission efficiency in ramping process are all given in this application.

The beam lifetime $\tau$ is defined as the inverse relative decay rate of the beam current. In the practical operation of a ring accelerator, the beam current exhibits an approximate exponential decay, it can be expressed by Eq. (2.9.1):
$$I(t) = I_0 exp(-t/\tau) \tag{2.9.1}$$
here, $\tau$ is the beam lifetime and $I_0$ is the beam current at t=0.

Taking logarithms on both sides of Eq. (2.9.1) and performing simple arithmetic, one can obtain
$$log(I_0)\tau - log(I(t))\tau = t \tag{2.9.2}$$
Assuming minimal variation in beam lifetime over a relatively short time period, the beam lifetime is $\tau_{1a}$ at the moment $t_1$ and $\tau_{1b}$ at other moments. According to Eq. (2.9.2) the following system of linear equations can be obtained:
$$\begin{pmatrix} log(I_0(t_1)) & log(I_0(t_1)) \\ log(I_0(t_1)) & log(I_0(t_2)) \\ log(I_0(t_1)) & log(I_0(t_3)) \\ log(I_0(t_1)) & log(I_0(t_4)) \end{pmatrix} \begin{pmatrix} \tau_{1a} \\ \tau_{1b} \end{pmatrix} = \begin{pmatrix} t_1 \\ t_2 \\ t_3 \\ t_4 \end{pmatrix} \tag{2.9.3}$$
The most approximate solution can be obtained by using the method of least squares:
$$\tilde{x} = (A^T A)^{-1} A^T \tag{2.9.4}$$
Eq. (2.9.4) can be solved to obtain $\tau_{1a}$ and $\tau_{1b}$.

The unreasonable data due to DCCT jitter is pre-processed so as to remove some points before calculation. The injection efficiency is obtained by calculating the slope of the charge of LBICT2 at the end of the transport line and the highest current at the moment of injection. The transmission efficiency is calculated by the DCCT current data before the ramping enabled signal and the DCCT current data after the power supply ramping table reach the required index. The GUI of this application is shown in Fig. 12.

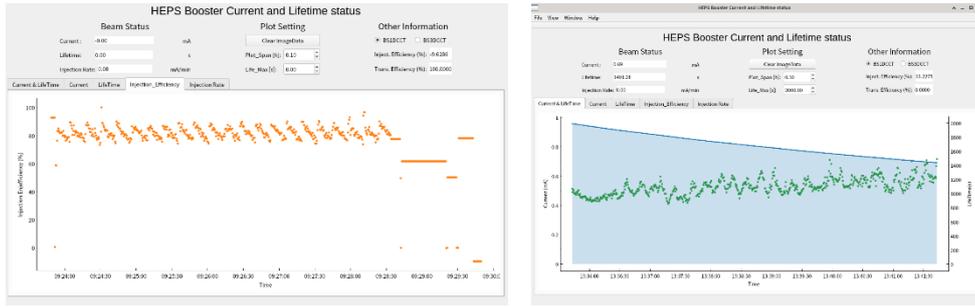

**Fig. 12.** The GUI of injection efficiency and beam lifetime measurement application

## 3. The application of HLAs during the beam commissioning

HLAs play a crucial role in the beam commissioning, and meanwhile, the beam commissioning is also a process of optimizing and improving HLAs.

Before the beginning of beam commissioning, only two ways of generating energy curves were considered in the booster controller application. One way is to directly keep the physical quantity unchanged based on the beam tuning results under the 500 MeV conditions, and generate the entire energy curve according to the energy scale. The other way is to optimize a certain energy point on the basis of a complete energy curve, and after optimization, make minimal changes to the energy ramp curve while ensuring that it does not exceed the power supplies' voltage limit. However, during the actual beam commissioning process, we found that due to the inconsistency between the power supply variation curve we used and the one used during magnetic field measurement, as well as the hysteresis effect, it was necessary to multiply the quadrupole magnets by different factors in different energy intervals to prevent significant beam loss during the energy ramping process. So, we added the function of the proportional adjustment of a magnet power supply ramping curve segment.

During the phase of improving the transmission efficiency, measurements and corrections were primarily conducted for the orbit and tune. Due to the unbinding of the reference frequency of booster and Linac, the RF fequency of the booster can only be operated at 499.8 MHz. However, this frequency was not matched with the actual circumference, resulting in a non-zero average orbit for the beam. Consequently, we needed to correct the orbit to a non-zero value. This issue was not expected before the beam commissioning. In order to better correct the orbit, we added an option in the global orbit correction program to set the target orbit to a non-zero value. In the tune adjusting application, we designed a function to measure the response of the quadrupole magnet strength to the tune. However, during actual beam commissioning, since the tune cannot be read online, the adjustment of the tune was based on theoretical response. Due to similar reason, in chromaticity measurement it also required manual input.

## 4. Summary

The HLAs server as an essential tool for beam commissioning. The HLAs for the HEPS booster have been developed based on *Pyapas*. The HLAs met the requirements initial commissioning of HEPS booster. During the beam tuning process, these HLAs were continuously optimized according to new requirements from the commissioning. Many apllications can be applied in the beam commissioning of the HEPS storage ring with minor modifications.

**Funding information**

The following funding is acknowledged: High Energy Photon Source (HEPS), a major national science and technology infrastructure in China; National Natural Science Foundation of China (grant Nos. 12005239, 11922512, 12275284)